\documentclass[aps,prl,reprint,nofootinbib,twocolumn,superscriptaddress,showpacs,showkeys,longbibliography]{revtex4-1}
\usepackage{eurosym}
\usepackage{amsmath,amssymb,amstext}
\usepackage[usenames,dvipsnames]{color}
\usepackage{graphicx}
\usepackage{braket}
\usepackage{natbib}
\usepackage{comment}
\usepackage{dcolumn}
\usepackage[english]{babel}
\usepackage{wasysym}
\usepackage[colorlinks,bookmarks=false,citecolor=blue,linkcolor=red,urlcolor=blue]{hyperref}

\newcommand{\new}[1]{\textcolor{black}{#1}}

\begin{document}
\title{  Dispersion and the  transport of  exciton-polaritons in an  optical conveyor belt}
\author{Xingran Xu}
\email{thoexxr@hotmail.com}
\affiliation{School of Science, Jiangnan University, Wuxi 214122, China}
\author{Chunyu Jia}
\affiliation{College of Physical Science and Technology, Bohai University, Jinzhou 121013, China}
\author{Xin-Xin Yang}
\affiliation{Shanghai Qizhi Institute and Shanghai Artificial Intelligence Laboratory, Xuhui District, Shanghai 200232, China}
\affiliation{\new{State Key Laboratory of Surface Physics, Institute of Nanoelectronics and Quantum Computing, Department of Physics, Fudan University, Shanghai 200438, China}}

\date{\today}

\begin{abstract}
The growing interest in exciton-polaritons has driven the need to manipulate their motion and engineer their band structures to the forefront of contemporary research. This study explores the band structures that emerge from a spatially modulated potential, ingeniously realized through the use of an optical conveyor belt. By leveraging Bloch theory and conducting a meticulous analysis of the time evolution of polariton intensity in Fourier space, we have derived the energy dispersion relations both analytically and numerically within the context of a static lattice model. For time-dependent potentials, we employ the Lagrange variational method to elucidate the dynamics of polariton motion. Our results reveal that polaritons exhibit linear dispersion and follow linear trajectories with minor oscillations superimposed. This investigation not only deepens our fundamental understanding of exciton-polaritons but also provides a robust tool for advancing photonic devices and exerting precise control over current transport in quantum computing. Our findings pave the way for future innovations in high-speed and high-performance technologies.
\end{abstract}
\maketitle

\section{I. Introduction}\label{section:one}

In the realm of quantum physics and condensed matter science, the study of exciton-polaritons has emerged as a vibrant and interdisciplinary field \cite{Rev1,Rev2,Rev3,Rev0}. Exciton-polaritons are quasiparticles that arise from the strong coupling between excitons (bound electron-hole pairs) in a semiconductor and photons in an optical cavity. This coupling results in a hybrid state of matter and light, combining the properties of both excitons and photons.

The unique characteristics of exciton-polaritons, including their low effective mass, adjustable nonlinear interactions \cite{Takemura_2014,PhysRevLett.122.047402}, and quantum coherence, have garnered significant attention for both fundamental research and practical applications. They are ideal platforms for realizing Bose-Einstein condensation (BEC) at room temperature \cite{Lerario_2017,Fraser_2017,Su_2020NP,K_dziora_2024}. Moreover, they provide a fertile ground for exploring non-Hermitian physics phenomena, such as exceptional points (EPs) \cite{Gao2018a,Nada2017,Khurgin:20,PhysRevResearch.6.013148}, where the eigenvalues and eigenvectors of a system coalesce, leading to intriguing behaviors such as unidirectional invisibility and enhanced sensitivity in sensing applications \cite{Chen_2017EP,Lee_2025EP}. Additionally, the non-Hermitian skin effect, where eigenstates accumulate at the boundaries of a system, can also be investigated in the context of exciton-polaritons \cite{mandal2021topological,Xu2021,XU2022,Xu_2025}. This effect has profound implications for the design of robust topological devices and the manipulation of light-matter interactions in a non-reciprocal manner \cite{huawen2021,Ruiqi2023,Chen_2024nhse}.

The manipulation of exciton-polariton dynamics via optical techniques is pivotal for unlocking their potential applications. In optics and photonics, the creation of optical lattices and the implementation of optical control can be achieved through various approaches, including external laser fields \cite{PhysRevX.10.011040,Caputo_2019}, spatial light modulators (SLMs) \cite{Benea_Chelmus_2021}, and optical conveyor belts \cite{del_Valle_Inclan_Redondo_2024,del_Valle_Inclan_Redondo_2023}. These techniques enable the engineering of external periodic potentials for both Hermitian and non-Hermitian lattices, allowing the formation of polaritons with distinctive band structures, such as flat bands \cite{Ge2018,Peotta2015,PhysRevLett.120.097401} and those exhibiting the zitterbewegung effect \cite{Sigur_sson_2024,Lovett_2023}.

Microcavity polariton solitons, excited on picosecond timescales, offer significant advantages for information processing over light-only solitons in semiconductor cavity lasers, which have nanosecond response times \cite{Sich_2011,Tanese_2013,Amo_2011}. The trajectories and equations of motion of the condensates can be obtained using the variational approach \cite{Xu_2018,Yilin2022,PhysRevB.111.165142,PhysRevA.111.023329}, providing a powerful means to investigate the dynamic behaviors of the condensates.

Recently, time-varying potentials have played a crucial role in shaping the dynamics and behavior of polariton systems. A time-varying potential can cause polaritons to accelerate or decelerate, depending on the shape and temporal variation of the potential \cite{PhysRevX.10.011040,Feres_2020}. Periodically modulated potentials can create periodic lattices that support the formation of polariton solitons \cite{PhysRevResearch.1.023030,ARKHIPOVA20232017}. Similarly, rotating time-varying potentials can induce the formation of vortices \cite{JIANG2022112368,Gnusov_2023}. Moreover, time-varying potentials can affect the coherence and phase stability of polariton condensates \cite{T_pfer_2020,Comaron_2025}.

\new{In the realm of polariton manipulation, the optical conveyor belt emerges as a distinctive method for transporting particles, setting itself apart from conventional techniques such as static optical lattices, gradient potentials, and spin-orbit coupling. Static optical lattices, for instance, offer a fixed periodic potential that is highly effective for trapping and manipulating particles in specific applications. However, they fall short in providing the dynamic control that the optical conveyor belt excels in, enabling precise and controlled transport of particles along a predefined trajectory \cite{Langbecker_2018}.
Gradient potentials, while capable of driving particles in a specific direction \cite{Song_2019GP}, often lack the fine-tuned control and flexibility inherent to the optical conveyor belt. This belt approach not only allows for high-precision manipulation of particles but also supports complex trajectories and adjustable transport parameters.
Spin-orbit coupling, another method, introduces a unique interaction between the spin and momentum of particles, leading to intriguing transport phenomena  \cite{Oue_2020}. Yet, its implementation is typically more complex and may not match the level of control offered by the optical conveyor belt.
The optical conveyor belt goes beyond simple directional control by enabling coherent transport of quantum states, a critical feature for quantum information processing. It can be dynamically adjusted to meet varying transport needs, offering unparalleled flexibility and adaptability. Furthermore, it minimizes the risk of decoherence and other quantum state degradation effects often seen in other techniques.
In summary, while static optical lattices, gradient potentials, and spin-orbit coupling each have their merits in polariton manipulation, the optical conveyor belt stands out due to its dynamic control, precision, and adaptability. These attributes make it a powerful tool for advanced applications in quantum physics and beyond.}

In this work, we investigate the dynamic behaviors and energy dispersion of wavepacket transport within an optical conveyor belt, a system that can be effectively modeled as a time-varying potential. Our aim is to uncover the intricate interplay between the wavepacket's motion and the dynamically changing potential landscape, revealing fundamental insights into the underlying physics and potential applications. We demonstrate that the dispersion relationship is closely connected with the transport behaviors, and we calculate the trajectories of the condensates both analytically and numerically.

\section{II. Model}\label{section:two}

\begin{figure}
 \centering
\includegraphics[width=\columnwidth]{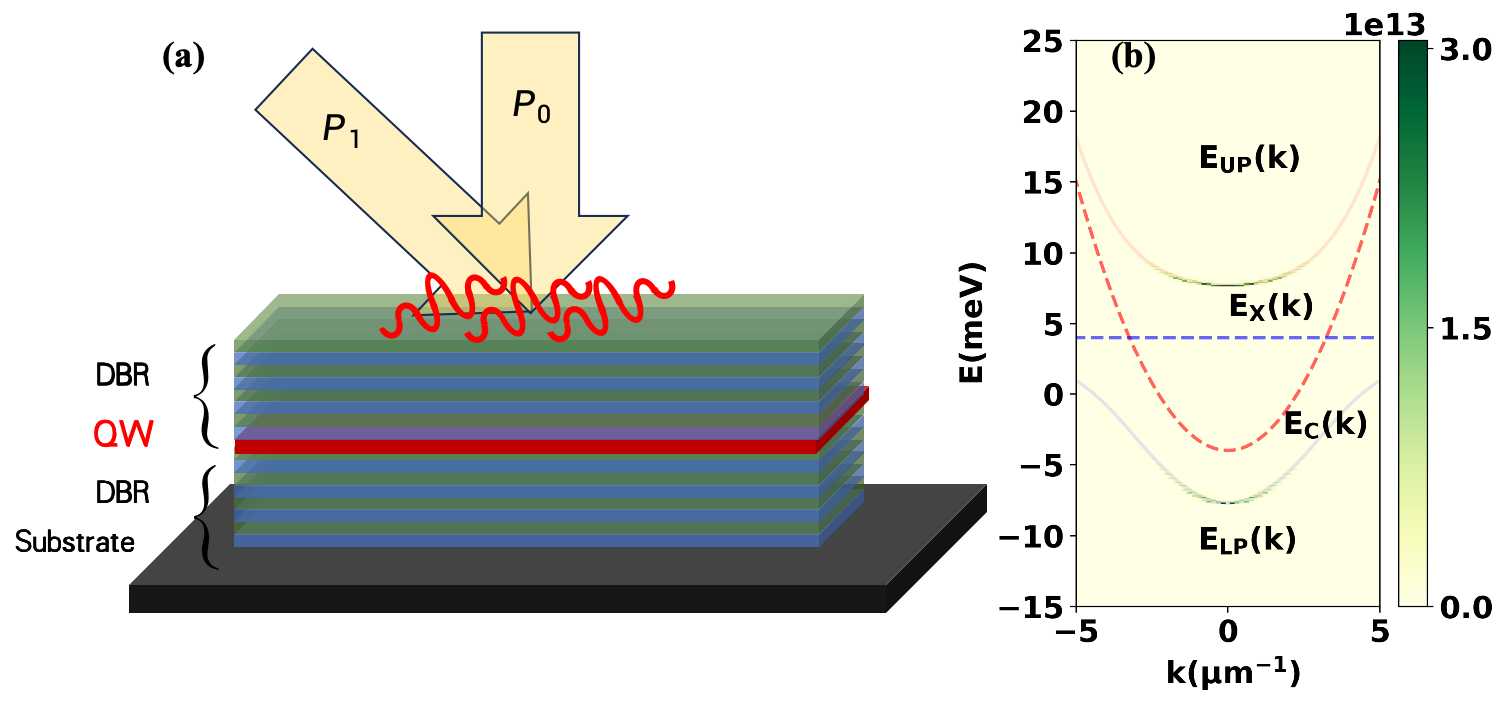}
\caption{  \new{(a) Sketch of the excitation scheme.
The quantum well (QW) is embedded within a planar micro-cavity formed by two distributed Bragg reflectors (DBRs). Two laser beams are injected: one drives the condensate formation; the second imprints an optical lattice (belt). (b) The dashed red and blue lines represent the kinetic energy of the cavity photons and excitons, respectively, with the expression $E_{c,\text{X}}(k)=\hbar^2k^2/2m_{c,\text{X}}$. The light red and blue lines indicate the analytical dispersion obtained from Eq. (\ref{LUbranch}), while the colorful figures are derived from the numerical simulation of Eqs. (\ref{Eqc})-(\ref{Eqex}). }}\label{F1}
\end{figure}

Exciton-polaritons are generally formed within a microcavity. This microcavity is composed of two parallel mirrors, namely distributed Bragg reflectors, with a thin layer of semiconductor material sandwiched in between. The strong coupling between the excitons in the semiconductor and the photons in the cavity gives rise to the formation of exciton-polaritons, as depicted in Fig. \ref{F1} (a). In our study, we focus on the polaritons that are formed through the strong coupling of excitons and photons. For the sake of simplicity, we describe these polaritons using the coupled Schrödinger equations \cite{Wurdack_2023,Song_2025,del_Valle_Inclan_Redondo_2024}:
\new{
\begin{eqnarray}
i\hbar\frac{\partial\psi_{c}}{\partial t}&=&\left[-\frac{\hbar^{2}}{2m_{c}}\frac{\partial^2 }{\partial_{x}^{2}}-\frac{\delta}{2}+i\frac{P_0-\hbar\gamma_c}{2}\right] \psi_c+\frac{\hbar\Omega}{2}\psi_{\text{X}},  \label{Eqc}\\
i\hbar\frac{\partial\psi_{\text{X}}}{\partial t}&=&\left[-\frac{\hbar^{2}}{2m_{\text{X}}}\frac{\partial^2}{\partial_{x}^{2}}+g|\psi_\text{X}|^2+\frac{\delta}{2}+V(t)\right]\psi_{ \text{X} }+\frac{\hbar\Omega}{2}\psi_{c}, \label{Eqex}
\end{eqnarray}}
where $\psi_c$ and $\psi_\text{X}$ represent the wavefunctions of cavity modes and the exciton modes.  $m_c$ and $m_\text{X}$ denote the mass of the cavity photons and excitons; $\delta$ and $\Omega$ are the two components' detuning and the Rabi splitting;\new{$\gamma_c$ is the decay of the photons and $P_0$ is the incoherent continuous-wave pump.} The time-dependent potential  $V(t)=V_{p}\left(1-\cos\left[\omega t-  G x\right]\right)$  acts as an optical conveyor belt, which can be dynamically modulated by introducing an additional pump laser within the exciton-polariton system.\new{ The angle offset between the lasers  $G$  determines the fringe periodicity, whereas the frequency offset  $\Delta f$  governs their speed and movement, with an angular frequency of  $\omega = 2\pi\Delta f$.
 The nonlinear interaction  $g$  of the polaritons, which originates from the exciton components, can be effectively tuned via the exciton-Feshbach resonance \cite{Takemura_2014,PhysRevLett.122.047402}.
}

\new{In our subsequent calculations, we set the parameters as follows: $m_c = 5 \times 10^{-5} m_0$ and $m_{\text{X}} = 0.1 m_0$, where $m_0$ represents the mass of the bare electron. The detuning is chosen to be $\delta = 8$ meV, and the Rabi splitting is $\hbar\Omega = 13.16$ meV. The significant Rabi splitting ensures robust coupling between the two components, thereby facilitating the formation of quasiparticles. The photon lifetime $\gamma_C^{-1}$ in GaN/AlGaN systems typically ranges from 0.3 to 1 ps. In our model, we set the incoherent pump strength $P_0$ near the threshold, with values spanning 1–4 $\mu$m$^{-2}$ps$^{-1}$ (equivalent to 10–50 W/cm$^{2}$) \cite{Carusotto2013,Byrnes_2014}. We focus on the condensation regime around the threshold, where $P_0 \approx \gamma_c$, and we ignore the effects of gain and loss in our theoretical model.}

The polaritons are formed by a pump laser $P_0$, and an additional laser $P_1$ is induced in the system, which can adjust the period and the motion of the polaritons. While the exact band in which a condensate(s) forms will depend on the periodicity of the lattice and the stability of the interference pattern, the band formation itself is robust and can be controlled through the angle between the two lasers ($G$) and the relative power of the angled beam ($P_1$) as shown in Fig. \ref{F1}(a) \cite{del_Valle_Inclan_Redondo_2024,Cheng:25,arxiv.2502.14986}. In the absence of a trapped potential, the dispersion relation for exciton-polaritons typically consists of two branches:
\begin{equation}  
E_{L,U}= \frac{1}{2}\left[E(k)  \pm\sqrt{(\delta -E_{c}(k)+E_{\text{X}}(k))^2+\hbar^2 \Omega ^2}\right], \label{LUbranch}
\end{equation}
 where  $E_{c,\text{X}}(k)=\hbar^2k^2/2m_{c,\text{X}}$ and $E(k)=E_{c}(k)+E_{\text{X}}(k)$. As shown in Fig. \ref{F1} (b), the dispersion for two particles splits into two branches. Both the upper and lower branches of the analytical dispersion for strong coupling of cavity photons and excitons correspond with the numerical results calculated from the coupling GP equations.   The colormap displays the density distribution of the quasiparticles, with the largest density of the two branches concentrated around zero momentum.

\section{III. Static band structures }
The depth and period of the optical lattice potential are of paramount importance in determining the localization of atoms. A deeper lattice potential exerts a stronger confining force on atoms, thereby reducing their kinetic energy and enhancing their stability. However, if the lattice is excessively deep, it may induce localization effects that trap polaritons within individual lattice sites, thereby diminishing coherence across the entire lattice structure. Similarly, the period of the potential also significantly impacts the localization of polaritons. In this section, we will further investigate the motion of polaritons under different lattice constants to gain a comprehensive understanding of these phenomena.
 
Given the translation symmetry of the periodic potential,  $V(x + 2\pi/G) = V(x)$ , the wavefunctions of excitons and cavity photons adhere to the principles of Bloch theory.
We can assume
\begin{eqnarray}
\psi_{c,\text{X}}(x)&=&e^{ikx} u_{c,\text{X}}(x), \\
V(x)&=&\sum_{n=-\infty}^{\infty}V_n e^{i n G x}
\end{eqnarray}  
 where $u_{c,\text{X}}(x)$ is a periodic function and the period is the same as the periodic potential. So we can use the Fourier transformation for the wavefunctions and the potentials. The static band structure can be obtained by 
 \begin{equation}
\hat{H}_{k}=\left[\begin{array}{cc}
\hat{H}_{c}-\frac{\delta}{2}\hat{I}_{N\times N} & \frac{\hbar\Omega}{2}\hat{I}_{N\times N}\\
\frac{\hbar\Omega}{2}\hat{I}_{N\times N} & \hat{H}_{\text{X}}+\hat{U}+\frac{\delta}{2}\hat{I}_{N\times N}
\end{array}\right], \label{Bloch}
 \end{equation} 
 where $\hat{I}_{N\times N} $ is the identity matrix, $\hat{H}_{c,\text{X}}$ is a diagonal matrix with the elements $\frac{\hbar^2 (k+nG)^2}{2m_{c,x}}$, and $\hat{U}=\text{diag}(\frac{V_0}{4},-1)_{N\times N}+\text{diag}(\frac{V_0}{4},1)_{N\times N}.$ We take $N$ an  odd number and $n$ from  $-(N -1)/2$ to $(N -1)/2$.
 
\begin{figure}[h]
\centering
\includegraphics[width=1\columnwidth]{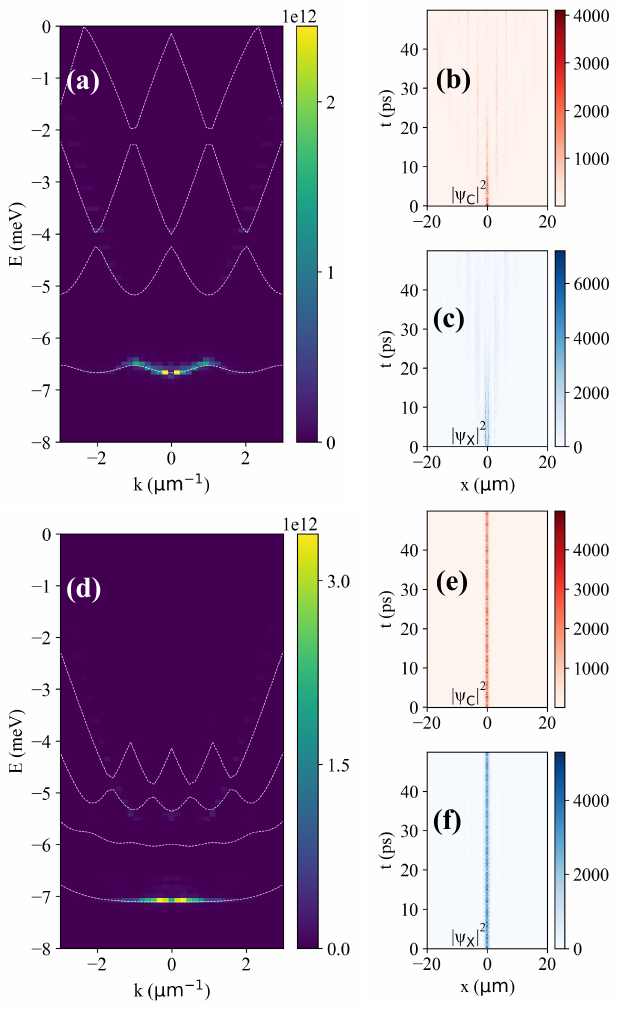}
\caption{The bandstructures for different lattice constants without interaction, determined through both analytical methods (white dashed lines) and numerical simulations (colorful figures), are presented in panels (a) and (d). The time evolution of the wavefunctions corresponding to the narrow potential is illustrated in panels (b) and (c), whereas the wavefunctions for the wide potential are displayed in panels (e) and (f). Parameters are: $V_p$= 10 (meV), $G$=2 ($\mu$m$^{-1}$) for (a)-(c) and $G$=1 ($\mu$m$^{-1}$)  for (e)-(f)}\label{F2}
\end{figure}

As depicted in Figs. \ref{F2} (a) and (d), the eigenvalues of Eq. (\ref{Bloch}) (represented by white dashed lines) align with the band structures obtained from the numerical simulation, with a constant energy shift. Since the majority of polaritons reside in the first energy band, our focus will be on the ground state of the polaritons.

In our numerical calculations, the initial state of the exciton is given by $\psi_{\text{X}} = \sqrt{N_0} e^{-x^2}$, which is a Gaussian wavepacket with a particle number $N_0$ of 5000. The center of this wavepacket is positioned at the midpoint of the lattice. Meanwhile, the initial state of the cavity photon $\psi_c$ is initialized as random noise.

\begin{figure*}
\centering
\includegraphics[width=2\columnwidth]{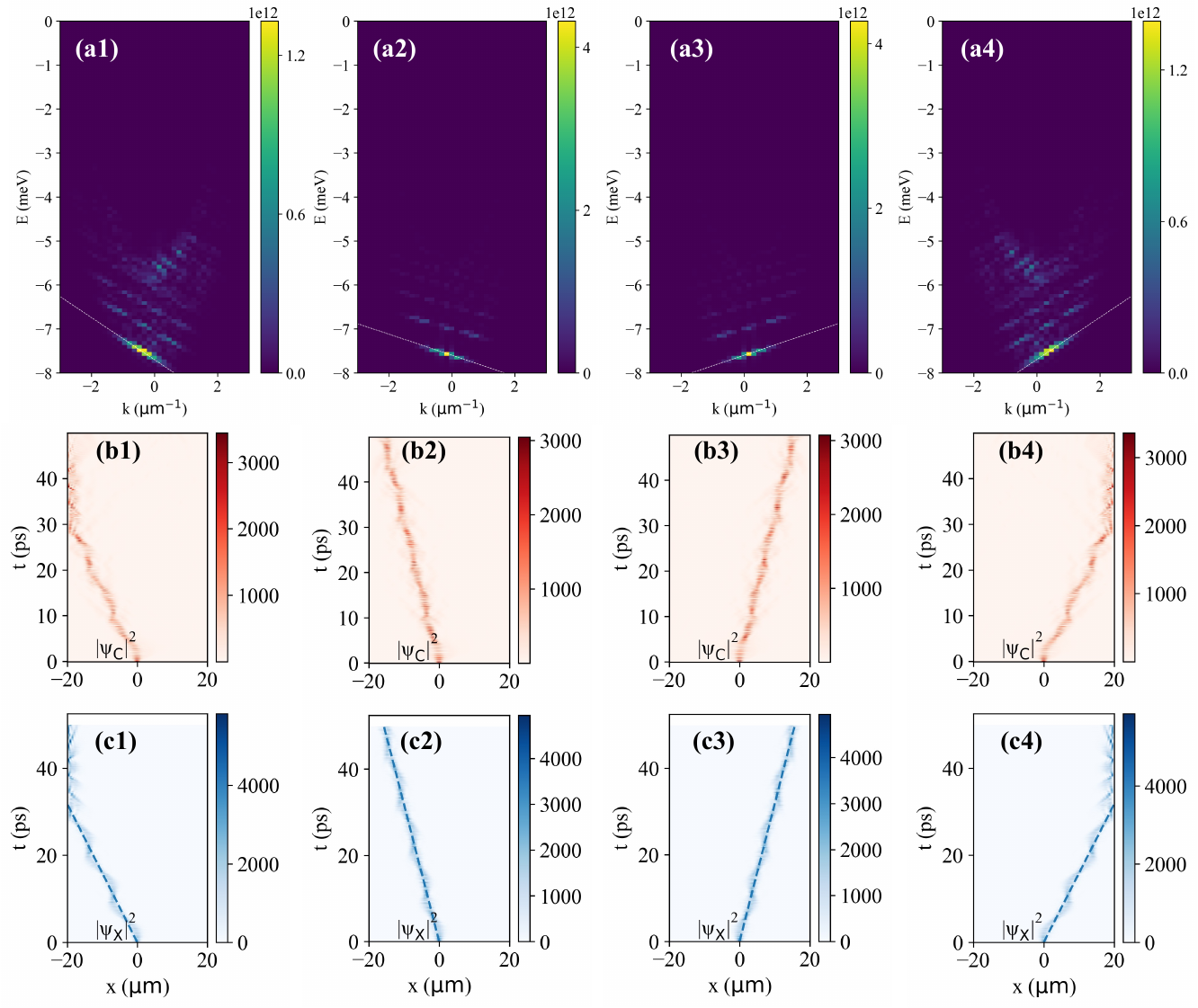}
\caption{ Intensity distribution (the first row) of polaritons in the energy-momentum space and the time evolution of the cavity mode (the second row) and the exciton mode (the third row). Parameters are: $V_p$= 10 (meV), $G$=0.2 ($\mu m^{-1}$), $g$=0, and $\Delta f$=-20, -10,10 and 20 (Ghz) for different columns.  }\label{F3}
\end{figure*}

\begin{figure*}
\centering
\includegraphics[width=1.8\columnwidth]{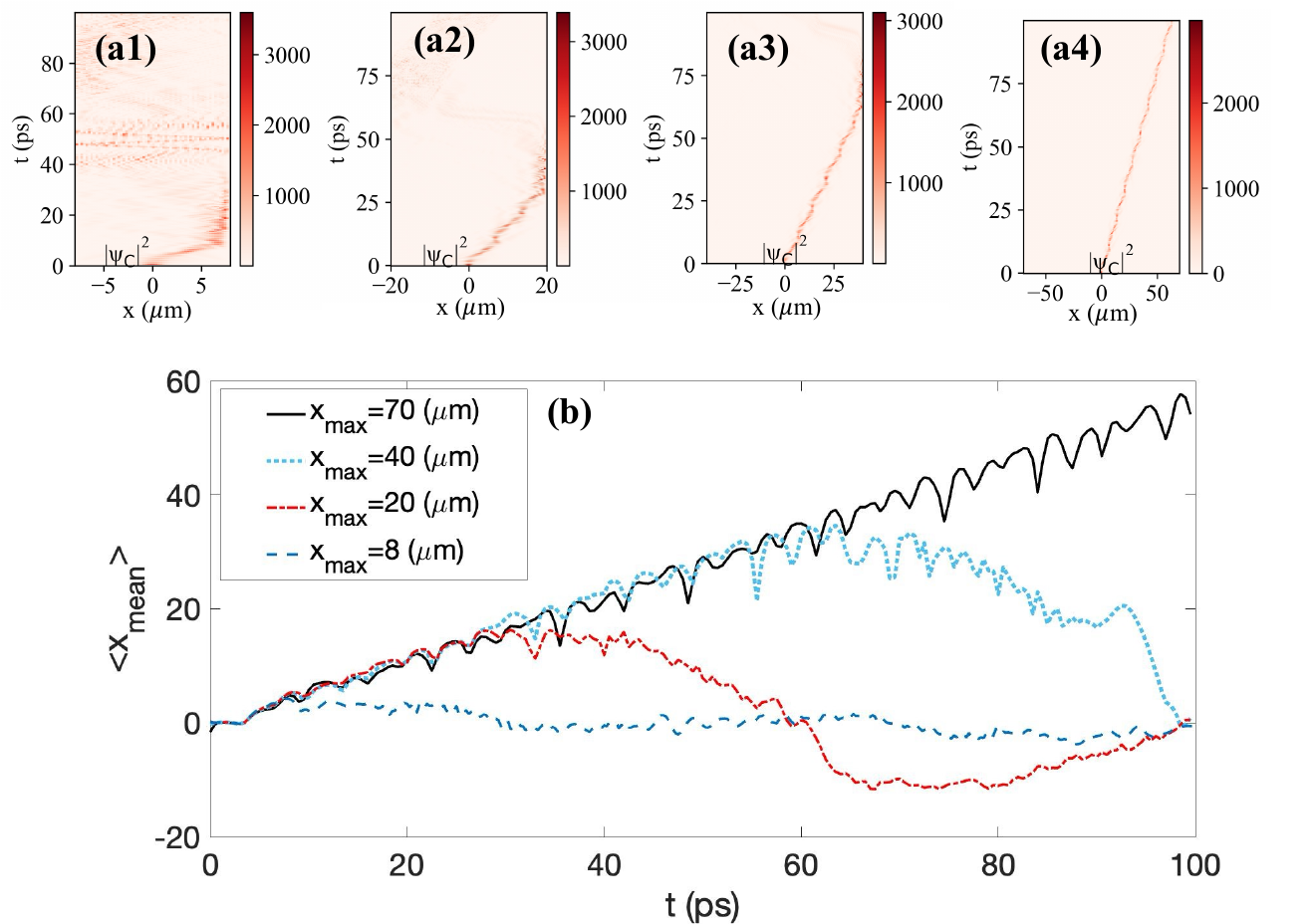}
\caption{ \new{Time evolution of wavepackets for cavity modes ((a1)-(a4))  across different system sizes, along with the mean position of the wavepackets for various sizes (b). Parameters are:  $V_p$= 10 (meV), $g$= 2$\times10^{-5}$ (meV/$\mu$m$^{-2}$),  $G$=0.2 ($\mu$m$^{-1}$),  $\Delta f$=20 (GHz), and the the system size $L=2x_{max}$= $16$, $40$, $80$, and $140$ ($\mu$m) for (a1)-(a4).  }}\label{sizeeffect}
\end{figure*}

When the lattice constant $G$ is small, the period of the potential, which is given by $2\pi/G$, becomes large. The wide potential provides polaritons with more time to propagate and extend to other sites, as shown in Figs. \ref{F2} (b) and (c). In contrast, when a narrow potential is produced, the polaritons get more restricted and are trapped in the initial lattice site \cite{Pickup_2020,1525069182844-335503387} shown in Figs. \ref{F2} (d) and (e).

In our following discussion, we are actively investigating the potential of dynamically modulated potentials to create an optical conveyor belt, which would enable precise control over the motion of polaritons. In this work, we have focused on addressing a critical challenge in this system: the potential collapse of polariton wavepackets. To mitigate this issue, we have delved into the study of stable trapped polaritons, seeking to develop robust mechanisms that can maintain their coherence and integrity throughout the transport process.

\section{IV. Dispersion and motion control}
When the frequency and the amplitude of the potential are nonzero, the polaritons will be pushed by this optical conveyor belt. Based on the above investigation, we can assume the steady wavefunctions for the cavity modes and exciton modes are Gaussian-like functions with the same central position.
\begin{equation}
\psi_{c,\text{X}}=\frac{1}{\left(\pi\sigma^{2}\right)^{1/4}}\exp\left[-\frac{\left(x-x_{0}\right)^{2}}{2\sigma^{2}}+i\kappa\left(x-x_{0}\right)+i\beta\left(x-x_{0}\right)^{2}\right],  \label{eq.ansa} 
\end{equation}
where  $x_{0}\left(t\right)$, $\kappa\left(t\right)$, $\sigma\left(t\right)$, and $\beta\left(t\right)$ are the variational parameters to be determined below. Specifically, $x_{0}\left(t\right)$ and $\sigma\left(t\right)$ are the center-of-mass position and width of the Gaussian wave function, respectively, $\kappa\left(t\right)$ is referred to as the wavenumber of the Gaussian function, $\beta\left(t\right)$ is related to the variation of the width. The time evolution of the variational parameters in Eq. (\ref{eq.ansa})  can be obtained via the Euler-Lagrangian equations \cite{Partanen_2019,He_2024} :
\begin{equation}
\frac{\partial L}{\partial q_{i}}-\frac{d}{dt}\left(\frac{\partial L}{\partial\dot{q_{i}}}\right)=0,  \label{eqL} 
\end{equation}
with $\dot{q_{i}}\equiv\frac{dq_{i}}{dt}$ and $q_{i}=x_{0}\left(t\right),\kappa\left(t\right),\sigma\left(t\right),\beta\left(t\right)$. In Eq. (\ref{eqL}), the Lagrangian $L=\int_{-\infty}^{+\infty}\mathcal{L}dx$ is referred to as the average Lagrangian of the coupling GP equation, where the Lagrangian density $\mathcal{L}$ is given in the Appendix. When we substitute the  Lagrangian into Eq. (\ref{eqL}) and obtain the equations of motion for the variational parameters $x_{0}\left(t\right)$, $\kappa\left(t\right)$:
\begin{eqnarray}
\frac{dx_{0}(t)}{dt}&=&\frac{\left(m_{c}+m_{\text{X}}\right)}{2m_{c}m_{\text{X}}}\hbar\kappa(t), \label{Dy1}\\ 
\frac{d\kappa(t)}{dt}&=&-\frac{V_{p}G}{2\hbar\sigma(t)^{2}}\exp\left[-\frac{G^{2}}{4\sigma(t)^{2}}\right]\textrm{sin}\left[Gx_{0}(t)-\omega t\right]. \label{Dy2}
\end{eqnarray}
Therefore, the velocity of polaritons is determined by the equation of $\kappa(t)$. The derivative of $\kappa$  is an exponential decay function times the periodically varying function. Therefore, we can assume $\kappa(t)=\kappa_0+\epsilon(t)$ and $|\epsilon(t)| \ll |\kappa_0|$. The central position of the wavefunctions will have uniform linear motion with oscillation with $<x_0(t)>=\kappa_0 t+A(t)\cos(\nu t+\phi)$. For a conveyor belt, the velocity of the linear motion $\kappa_0=\omega/G$ and the oscillation have a relation with the width of the condensation $\sigma(t)$.  

The dispersion of a moving wave packet in an optical conveyor belt exhibits a linear relationship between energy and momentum, as depicted in Figs. \ref{F3} (a1)-(a4). The white dashed lines represent the relation  $E(k) = \frac{\hbar^2 \kappa_0}{2 m_\text{eff}} k$ , where  $m_\text{eff} = \frac{m_{c} m_{X}}{m_{c} + m_{e}}$ . This theoretical relation aligns well with the results obtained from numerical simulations. The sign of the frequency  $\omega$  governs both the direction of polariton transport and the tilt direction of the dispersion. When  $\omega < 0$ , the polaritons move to the left, and the dispersion tilts to the left, as illustrated in the first and second columns of Fig. \ref{F3}. Conversely, when  $\omega > 0$ , the polaritons move to the right, and the dispersion tilts to the right, as shown in the third and fourth columns of Fig. \ref{F3}.

\new{When the wave packet reaches the boundary, the polaritons will oscillate along the boundary and eventually collapse into the background. This behavior is dictated by the open boundary condition. However, if the conveyor speed $\omega/G$ is significantly slower, the propagation distance along the boundary will be substantially extended.
As is shown in Fig. \ref{sizeeffect}, we plot the time evolution of the wavepackets with different system sizes. Even though they have the same initial states and off-resonant frequency, the lifetime of the wavepackets is different. The wavepackets in the bulk can propagate without obstacles, as is shown in Figs. \ref{sizeeffect} (a1)-(a4). However, when they reach the boundary, the wavepackets will bounce along the edge due to the open boundary condition. If the system size is large enough, the wavepackets can survive much longer, as is shown in Fig. \ref{sizeeffect} (a4).
The mean position of the wavepackets of the photon modes can be calculated by:
\begin{equation}
<x_{\text{mean}}>=\frac{\int \psi_c^*(x,t)|\hat{x}|\psi_c(x,t) dx}{\int |\psi_c(x,t)|^2 dx}.
\end{equation}
The mean position of the wavepackets of the larger system size (140 $\mu$m) can maintain linear propagation with minor oscillation for a much longer time than that of the smaller system size, as shown in Fig. \ref{sizeeffect} (c).  }

The frequency of the varying potential exerts a significant influence on their behavior and characteristics. Firstly, the sign of the frequency determines the direction of polariton transport. When the frequency is negative, polaritons move to the left. Conversely, when the frequency is positive, they move to the right. This directional control is crucial for applications that require precise manipulation of polariton motion. Secondly, the magnitude of the frequency affects the velocity of the polaritons. A higher frequency results in a faster movement, while a lower frequency leads to slower motion. This allows for adjustable speed control, which can be tailored to specific experimental or technological requirements.

\new{\section{V. The stability of the moving wavepackets }
In this section, we will delve into the stability of wavepackets under varying potential parameters and the interaction strengths. Our previous calculations have been based on the assumption of weak exciton interactions and a small off-resonance frequency for the two laser pumps. However, it is important to recognize that experimental parameters can span a much broader range. Consequently, we will explore how these diverse parameters influence the stability of the wavepackets.}

\begin{figure}[h]
\centering
\includegraphics[width=1\columnwidth]{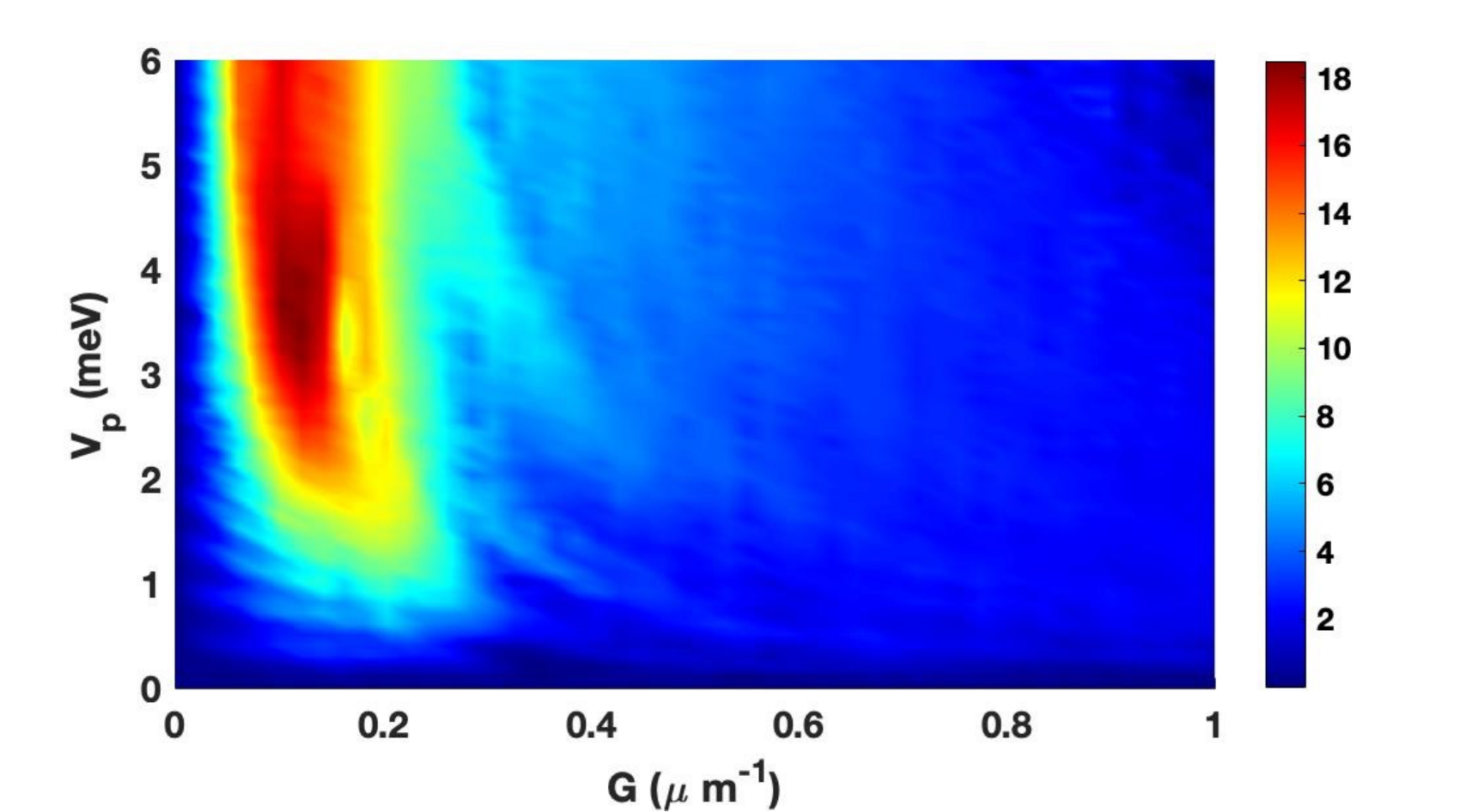}
\caption{ \new{The central position of the wavepackets at 20 ps with different potential depths and the lattice constant.  Parameters are:  $g$= 2$\times10^{-5}$ (meV/$\mu$m$^{-2}$),  and  $\Delta f$=20 (GHz) for different columns. } }\label{stable}
\end{figure}

The potential parameters are pivotal in determining the characteristics of wavepackets. Alterations in these parameters can profoundly influence both the spatial and temporal attributes of the wavepackets. For example, modifying the depth or shape of the potential well can significantly impact the confinement and dispersion of the wavepackets. As illustrated in Fig. \ref{F2}, a deeper potential well tends to result in more localized wavepackets, whereas a shallower well may lead to increased dispersion and potential instability. 
\new{We present the central position of the wavepackets in Fig. \ref{stable}. The potential depth should significantly exceed the linear transport kinetic energy, given by $\frac{\hbar^2}{2m_c}(\frac{\omega}{G})^2$. The ansatz provided in Eq. (\ref{eq.ansa}) dictates the width of the wavepacket and is contingent upon the lattice constant taking on specific values. This relationship underscores the intricate interplay between the potential landscape and the wavepacket dynamics, highlighting the importance of carefully tuning potential parameters to achieve desired wavepacket behavior. }

Moreover, the off-resonance frequency of the laser pumps is another critical parameter that can substantially affect the stability and coherence of the wavepackets. While our initial calculations have focused on small off-resonance frequencies, larger deviations from resonance can have notable impacts. A large off-resonance frequency can reduce the coupling between the laser pumps and the exciton system, thereby affecting the coherence and stability of the wavepackets. Conversely, fine-tuning the off-resonance frequency to specific values can optimize the coupling and potentially enhance the stability of the wavepackets.

\begin{figure}[h]
\centering
\includegraphics[width=1\columnwidth]{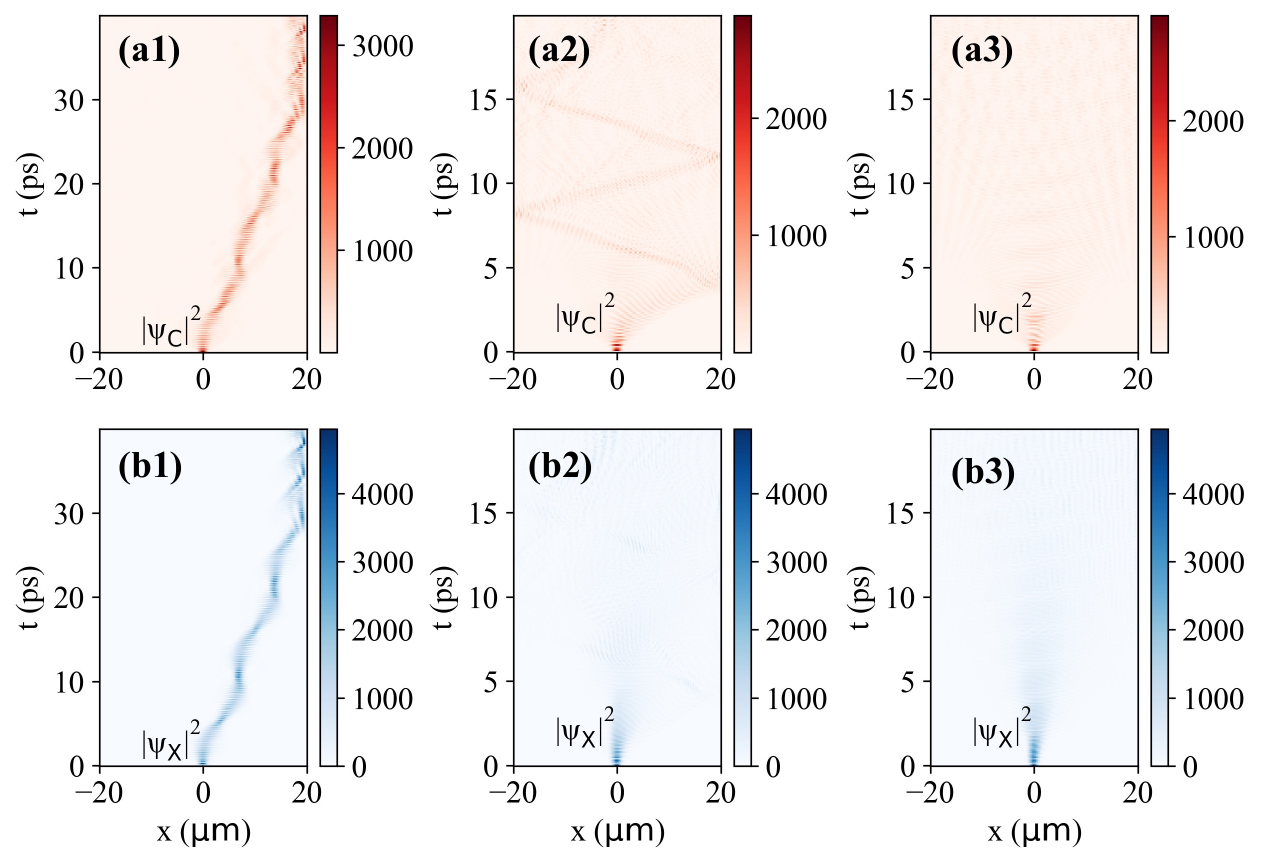}
\caption{ Time evolution of the cavity modes (the first row) and exciton modes (the second row) with different potential engineering. Parameters are:  $V_p$= 10 (meV), $g$= 2$\times10^{-5}$ (meV/$\mu$m$^{-2}$)  ,$G$=0.2 ($\mu$m$^{-1}$), and  $\Delta f$=20, 200, 500 (GHz) for different columns.  }\label{freq}
\end{figure}

As depicted in Fig. \ref{freq}, we present the temporal evolution of polaritons subjected to potentials with varying frequencies. When the frequency is significantly lower, the polaritons exhibit a slower movement and are much more stable, a phenomenon also observable in Fig. \ref{freq}. Conversely, when the frequency of the polaritons becomes substantially higher, the wavepackets tend to dissipate into the background, as illustrated in Figs. \ref{freq} (a3) and (b3).

\new{Moreover, the interaction strength between excitons is another key factor. While we have initially considered weak interactions, stronger interactions can introduce nonlinear effects that may either stabilize or destabilize the wavepackets. For example, attractive interactions ($g\le 0$)can lead to the formation of bound states or condensates, which may enhance the stability of the wavepackets. On the other hand, repulsive interactions can cause the wavepackets to spread out, potentially leading to instability.}
\begin{figure}[h]
\centering
\includegraphics[width=1\columnwidth]{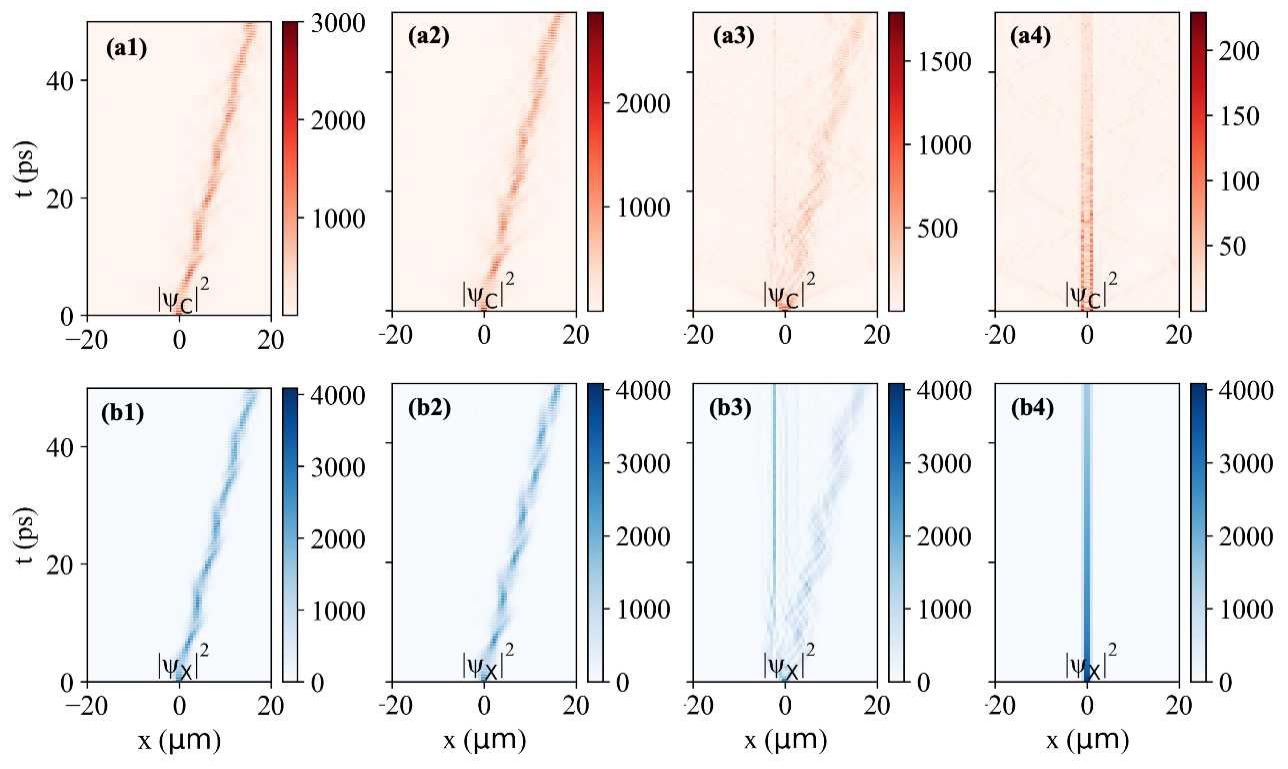}
\caption{ \new{Time evolution of the cavity modes (the first row) and exciton modes (the second row) with different interaction strength. Parameters are:  $V_p$= 10 (meV), $G$=0.2 ($\mu$m$^{-1}$), $\Delta f$=10 (GHz), and $g$=2$\times10^{-5}$, 2$\times10^{-4}$, 2$\times10^{-3}$ and 2$\times10^{-2}$ (meV/$\mu$m${^-2}$)   for different columns.}  }\label{interaction}
\end{figure}
\new{As illustrated in Fig. \ref{interaction}, we have systematically increased the interaction strength by four orders of magnitude to thoroughly examine the stability of the initial wavepacket. During this process, the corresponding interaction energy  $g|\psi_\text{X}|^2$  varies significantly, ranging from 0.08 meV to 80 meV. When the interaction strength is relatively small, the wavepacket remains stable and exhibits linear transport behavior, as clearly demonstrated in the first and second columns of Fig. \ref{interaction}. However, as the interaction strength $g$  increases, distinct changes occur. Specifically, when the interaction strength  $g$  reaches intermediate values, a portion of the wavepacket becomes trapped in the middle of the lattice sites and no longer moves with the optical conveyor belt, as illustrated in Figs. \ref{interaction} (a3) and (b3). When the interaction strength becomes sufficiently large, the wavepacket loses its linear transport property entirely and remains stationary in the middle of the lattice, localized at more than one site, as shown in Figs. \ref{interaction} (a4) and (b4). This transition underscores the pivotal role of interaction strength in determining the transport and stability characteristics of the wavepacket. }

\new{In summary, the stability of wavepackets is highly sensitive to the potential parameters, interaction strengths, and off-resonance frequencies of the laser pumps. By systematically exploring these parameters, we aim to gain a comprehensive understanding of the conditions under which stable wavepackets can be achieved. This analysis will not only provide valuable insights for theoretical studies but also guide experimental efforts in optimizing the stability of wavepackets for various applications.}

\section{VI. Discussion}
In this study, we have conducted comprehensive numerical simulations and provided in-depth analytical insights into the dispersion and trajectories of exciton-polaritons within an optical conveyor belt. Specifically, when investigating a single wavepacket confined in a time-varying potential, we have observed a notable transition in the dispersion of exciton-polaritons from the conventional Bloch relation to a linear relation. Concurrently, the trajectories of these wavepackets display linear propagation, albeit with minor oscillations superimposed.

\new{The optical conveyor belt technique has shown significant promise in the manipulation and transport of polaritons, offering distinct advantages over other methods such as static optical lattices, gradient potentials, and spin-orbit coupling. Recent studies have demonstrated that optical conveyor belts can achieve high transport efficiency and speed while minimizing heating and loss mechanisms.This approach allows for non-reciprocal band structures and the possibility of modulation faster than the polariton lifetime, highlighting its potential for advanced quantum systems \cite{del_Valle_Inclan_Redondo_2024,Cheng_2025,Gao_2025}. The unique advantages of the optical conveyor belt include precise control over particle movement, adaptability to various particle types and experimental conditions, and the ability to follow complex trajectories. These features make it particularly suitable for applications requiring high-density transport and detailed spatial manipulation \cite{Wang_2016,Xu_2022OL} }

The proposed scheme for controlling polariton trajectories holds immense promise for propelling the development of the next generation of high-speed and high-performance technologies. As research progresses, we anticipate further refinements and innovations that will unlock even greater potential. For example, integrating our scheme with emerging materials and nanostructures could give rise to novel functionalities and applications that are currently unattainable. This advancement has the potential to revolutionize a wide range of fields, from optical communication to quantum computing, thereby heralding a new era of high-performance technologies.

\new{ \section{VII. DATA AVAILABILITY}  The data are available upon reasonable request from the authors.}

\section{VIII. Acknowledgements}
The work is supported by the National Natural Science Foundation of China (Grant No. 12404362) and the Fundamental Research Funds for the Central Universities (Grant No. JUSRP123027).

\bibliography{mybib}
 
\appendix
\begin{widetext}

\appendix

\section{S1. The  Lagrangian variational approach of the motion of the exciton-polariton with optical conveyor belt }

In this section, we develop an analytical study of the soliton motion using the Lagrangian variational approach.
According to the coupling Gross-Pitaevskii (GP)  equation,i.e.,
\begin{eqnarray}
i\hbar\frac{\partial\psi_{c}}{\partial t}&=&\left[-\frac{\hbar^{2}}{2m_{c}}\frac{\partial}{\partial_{x}^{2}}-\frac{\delta}{2}\right]\psi_{c}+\frac{\hbar\Omega}{2}\psi_{\text{X}},  \label{eq.1} \\
i\hbar\frac{\partial\psi_{\text{X}}}{\partial t}&=&\left[-\frac{\hbar^{2}}{2m_{\text{X}}}\frac{\partial}{\partial_{x}^{2}}+\frac{\delta}{2}+V\left(t\right)+g\left|\psi_{\text{X}}\right|^{2}\right]\psi_{\text{X}}+\frac{\hbar\Omega}{2}\psi_{c}. \label{eq.2} 
\end{eqnarray}
with the pumping potential is explicitly given by $V\left(t\right)=V_{p}\left(1-\cos\left[\omega t-Gx\right]\right)$.
To seek the solutions of the above GP equation, we use the Lagrangian variational approach. We assume a Gaussian function, i.e.,
\begin{equation}
\psi_{c,\text{X}}=\frac{1}{\left(\pi\sigma\left(t\right)^{2}\right)^{1/4}}\exp\left[-\frac{\left(x-x_{0}\left(t\right)\right)^{2}}{2\sigma\left(t\right)^{2}}+i\kappa\left(t\right)\left(x-x_{0}\left(t\right)\right)+i\beta\left(t\right)\left(x-x_{0}\left(t\right)\right)^{2}\right],  \label{eq.3} 
\end{equation}
where  $x_{0}\left(t\right)$, $\kappa\left(t\right)$, $\sigma\left(t\right)$, and $\beta\left(t\right)$ are the variational parameters to be determined below. Specifically, $x_{0}\left(t\right)$ and $\sigma\left(t\right)$ are the center-of-mass position and width of the Gaussian wave function, respectively, $\kappa\left(t\right)$ is referred to as the wavenumber of the Gaussian function, $\beta\left(t\right)$ is related to the variation of the width . 
The time evolution of the variational parameters in Eq. (\ref{eq.3})  can be obtained via the Euler-Lagrangian equations.  Thus the Euler-Lagrange equations for the variational parameters are given by 
\begin{equation}
\frac{\partial L}{\partial q_{i}}-\frac{d}{dt}\left(\frac{\partial L}{\partial\dot{q_{i}}}\right)=0,  \label{eq.4} 
\end{equation}
with $\dot{q_{i}}\equiv\frac{dq_{i}}{dt}$ and $q_{i}=x_{0}\left(t\right),\kappa\left(t\right),\sigma\left(t\right),\beta\left(t\right)$.

In Eq. (\ref{eq.4}) , the Lagrangian $L=\int_{-\infty}^{+\infty}\mathcal{L}dx$ is referred to as the average Lagrangian of the coupling GP equation, where the Lagrangian density $\mathcal{L}$ is given by
\begin{equation}
\mathcal{L}=\mathcal{L}_{c}+\mathcal{L}_{\text{X}}+\mathcal{L}_{c\text{X}},  \label{eq.5} 
\end{equation}
where we have
\begin{subequations}
\begin{eqnarray}
\mathcal{L}_{c} & = &\frac{i\hbar}{2}\left(\psi_{c}^{*}\partial_{t}\psi_{c}-\psi_{c}\partial_{t}\psi_{c}^{*}\right)-\frac{\hbar^{2}}{2m_{c}}\left|\nabla\psi_{c}\right|^{2}+\frac{\delta}{2}\left|\psi_{c}\right|^{2}, \\
\mathcal{L}_{\text{X}} & = &\frac{i\hbar}{2}\left(\psi_{\text{X}}^{*}\partial_{t}\psi_{\text{X}}-\psi_{\text{X}}\partial_{t}\psi_{\text{X}}^{*}\right)-\frac{\hbar^{2}}{2m_{\text{X}}}\left|\nabla\psi_{\text{X}}\right|^{2}-\left[\frac{\delta}{2}+V\left(t\right)+\frac{g}{2}\left|\psi_{\text{X}}\right|^{2}\right]\left|\psi_{\text{X}}\right|^{2},\\
\mathcal{L}_{c\text{X}} & =& -\frac{\hbar\Omega}{2}\psi_{\text{X}}\psi_{c}^{*}-\frac{\hbar\Omega}{2}\psi_{c}\psi_{\text{X}}^{*}. 
\end{eqnarray}
\end{subequations}
Inserting the ansatz  (\ref{eq.3})  into Eq. (\ref{eq.5}), we calculate the average Lagrangian $L$ in Eq. (\ref{eq.4}) as
\begin{eqnarray}
L &=& 2\hbar\kappa\left(t\right)\frac{dx_{0}\left(t\right)}{dt}-\hbar\sigma\left(t\right)^{2}\frac{d\beta\left(t\right)}{dt}-\frac{\hbar^{2}}{4m_{\text{eff}}\sigma\left(t\right)^{2}}\left(1+2\kappa\left(t\right)^{2}\sigma\left(t\right)^{2}+4\beta\left(t\right)^{2}\sigma\left(t\right)^{4}\right)  \nonumber \\
 &-&\frac{g}{2\sqrt{2\pi}\sigma\left(t\right)}-\frac{V_{p}}{\sigma\left(t\right)^{2}}\left(1-\exp\left[-\frac{G^{2}}{4\sigma\left(t\right)^{2}}\right]\cos\left[Gx_{0}\left(t\right)-\omega t\right]\right)-\hbar\Omega,  \label{Lagrangian} 
\end{eqnarray}
where $m_{\text{eff}}\equiv\frac{m_{c}m_{\text{X}}}{m_{c}+m_{\text{X}}}$.

By substituting Eq. (\ref{Lagrangian}) into Eq.  (\ref{eq.4}), we obtain the equations of motion for the variational parameters $x_{0}\left(t\right)$, $\kappa\left(t\right)$, $\sigma\left(t\right)$, and $\beta\left(t\right)$ in  Eq. (\ref{eq.3}) as
\begin{subequations}
\begin{eqnarray}
\frac{dx_{0}\left(t\right)}{dt}&=&\frac{\hbar}{2m_{\text{eff}}}\kappa\left(t\right), \label{Dy1}\\ 
\frac{d\kappa\left(t\right)}{dt}&=&-\frac{V_{p}G}{2\hbar\sigma\left(t\right)^{2}}\exp\left[-\frac{G^{2}}{4\sigma\left(t\right)^{2}}\right]\textrm{sin}\left[Gx_{0}\left(t\right)-\omega t\right], \label{Dy2}\\ 
\frac{d\beta\left(t\right)}{dt}&=&-\frac{\hbar\left(-1+4\beta\left(t\right)^{2}\sigma\left(t\right)^{4}\right) }{4m_{\text{eff}}\sigma\left(t\right)^{4}} 
+\frac{g}{4\sqrt{2\pi}\hbar\sigma\left(t\right)^{3}} \nonumber \\
&+&\frac{V_{p}}{4\hbar\sigma\left(t\right)^{6}}\left(4\sigma\left(t\right)^{2}+\exp\left[-\frac{G^{2}}{4\sigma\left(t\right)^{2}}\right]\cos\left[Gx_{0}\left(t\right)-\omega t\right]\left(G^{2}-4\sigma\left(t\right)^{2}\right)\right), \label{Dy3}\\ 
\frac{d\sigma\left(t\right)}{dt}&=&\frac{\hbar}{m_{\text{eff}}}\beta\left(t\right)\sigma\left(t\right). \label{Dy4}
\end{eqnarray}
\end{subequations}
Eq. (\ref{Dy1})--(\ref{Dy4}) are the key results of this work, which describe the time evolution of the variational parameters in Eq. (\ref{eq.3}).

\end{widetext}
\end{document}